\documentclass[aps,pra,reprint,
twocolumn,amsmath,superscriptaddress,amssymb,showpacs,nofootinbib]{revtex4-1}
\usepackage{amsmath,braket,amssymb}
\usepackage[final]{graphicx}
\usepackage{subfigure}
\usepackage[english]{babel}
\usepackage[utf8]{inputenc}
\usepackage{mathtools}
\usepackage{empheq}
\usepackage{tensor}
\usepackage{cancel}

\usepackage{simplewick}
\usepackage{xfrac}

\usepackage[usenames,dvipsnames]{color}

\newcommand{\blue}{\color{blue}}

\usepackage{enumitem}
\usepackage{slashed}
\usepackage{graphicx}
\usepackage{xcolor}
\usepackage{tikz}

\usepackage{bm}
\usepackage[colorlinks=true,citecolor=blue,linkcolor=blue,urlcolor=blue]{hyperref}
\usepackage{dsfont}
\usepackage{changepage}
\usepackage{array}
\usepackage{soul}
\usepackage[capitalise]{cleveref}
\crefname{section}{Sec.}{Secs.}
\Crefname{section}{Sec.}{Secs.}
\usepackage{xifthen}
\usepackage{xargs}

\usepackage{bm}
\renewcommand{\vec}[1]{\boldsymbol{\mathbf{#1}}}

\graphicspath{{./}{./figures/}}

\newcommand{\bit}{\begin{itemize}}
\newcommand{\eit}{\end{itemize}}

\newcommand{\f}{\frac}
\renewcommand{\>}{\right\rangle}
\newcommand{\<}{\left\langle}
\newcommand{\ba}{\begin{align}}
\newcommand{\ea}{\end{align}}
\newcommand{\be}{\begin{equation}}
\newcommand{\ee}{\end{equation}}
\newcommand{\bi}{\begin{itemize}}
\newcommand{\ei}{\end{itemize}}
\newcommand{\lf}{\left(}
\newcommand{\ri}{\right)}
\newcommand{\dd}{\mathrm{d}}

\newcommand{\Tr}{\operatorname{Tr}}

\DeclareMathAlphabet{\mymathbb}{U}{BOONDOX-ds}{m}{n}

\usepackage{comment}

\usepackage{braket}
\usepackage[normalem]{ulem}

\newcommand{\nocontentsline}[3]{}
\newcommand{\tocless}[2]{\vspace{2em}\bgroup\let\addcontentsline=\nocontentsline#1{#2}\egroup}

\begin{document}

\title{Continuous symmetry breaking in 1D spin chains and 1+1D field theory}
\author{Adam Nahum}
\affiliation{Laboratoire de Physique de l’\'Ecole Normale Sup\'erieure, CNRS, ENS \& Universit\'e PSL, Sorbonne Universit\'e, Universit\'e Paris Cit\'e, 75005 Paris, France}

\date{\today}

\begin{abstract}
We argue that ground states of 1D spin chains can spontaneously break U(1) ``easy-plane''  spin rotation symmetry, via  true long-range order of $(S^x, S^y)$, at the phase transition between  two quasi-long-range-ordered phases. The critical point can be reached by tuning a single parameter in a Hamiltonian with the same symmetry as the XXZ model, without further fine-tuning. Equivalently, it can arise in systems of bosons with particle-hole symmetry, as a long-range-ordered transition point between two quasi-long-range-ordered superfluids. Our approach is to start with the continuum field theory of the isotropic Heisenberg ferromagnet and consider generic perturbations that respect easy-plane symmetry. We argue for a renormalization-group flow to a critical point where long-range order in $(S^x, S^y)$ is enabled by coexisting critical fluctuations of $S^z$. (We also discuss multicritical points where further parameters are tuned to zero.) These results show that it is much easier to break continuous symmetries in 1D than standard lore would suggest. The failure of standard intuition for 1D chains (based on the quantum--classical correspondence) can be attributed to Berry phases, which  prevent the 1+1D system from mapping to a classical 2D spin model. The present theory  also gives an example of an ordered state whose Goldstone mode is interacting even in the infra-red, rather than becoming a free field. 
\end{abstract}

\maketitle

\section{Introduction}

It is commonly assumed that long-range order (LRO) which breaks a continuous symmetry is impossible in ground states of 1D quantum chains, or rather, that it is impossible in cases where the order parameter has nontrivial quantum fluctuations  (see below).
This lore is based partly on the 
the correspondence between quantum statistical mechanics in 1D and classical statistical mechanics in 2D,
together with Hohenberg-Mermin-Wagner-like theorems  \cite{hohenberg1967existence,
mermin1966absence,
mermin1967absence,
coleman1973there,
mcbryan1977decay,
garrison1972absence,
klein1981absence,
frohlich1981absence,halperin2019hohenberg}
ruling out continuous symmetry breaking at nonzero temperature in 2D. 
For example, it has been believed that a 1D superfluid can at most have quasi-long-range order \cite{halperin2019hohenberg}. 
There are also rigorous theorems, based on the uncertainty principle, ruling out  symmetry breaking in 1D quantum ground states under the assumption that the corresponding conserved charge has a finite susceptibility \cite{pitaevskii1991uncertainty,shastry1992bounds,momoi1996quantum}. This approach gives for example a  proof of the absence of LRO in the antiferromagnetic Heisenberg  chain \cite{shastry1992bounds}.

Nevertheless there is no completely general theorem forbidding continuous symmetry breaking in 1D, and 
indeed it is clear from known examples that there cannot be.
The simplest example of 1D LRO is the isotropic Heisenberg ferromagnet, 
${\mathcal{H}=-J\sum_i\vec S_i. \vec S_{i+1}}$.
The ground states of this model are superpositions of 
trivial polarized product states such as $\ket{\ldots \uparrow\uparrow\uparrow\uparrow \ldots}$,
so that the correlator 
$\< \vec S_i. \vec S_j\>$
is long-ranged at zero temperature.\footnote{The theorems of Refs.~\cite{pitaevskii1991uncertainty,momoi1996quantum} do not apply because the appropriate susceptibility is not finite. SU(2) ferromagnetism also arises  for the eta-pairing SU(2) symmetry in a variant of the Hubbard model \cite{essler1992new}.} 
This example is usually set aside as a special case, 
since  the order parameter $\sum_i \vec S_i$ commutes with the Hamiltonian, allowing for these fully polarized ground states with no quantum fluctuations.
(While the isotropic ferromagnet could be regarded as a somewhat trivial limit, we will find that it is a useful starting point for locating nontrivial examples of LRO.)

More recent examples show that LRO is possible even when the order parameter does not commute with the Hamiltonian \cite{watanabe2024critical,fava2024heisenberg}, though these examples retain 
more structure than a generic model with minimal [i.e. U(1)] continuous symmetry.
Refs.~\cite{watanabe2024critical,ogunnaike2023unifying} described an interesting series of  ``frustration free'' spin chains  that break easy-plane U(1) symmetry.  The additional structure in these models, which was suggested as the key property allowing LRO \cite{watanabe2024critical}, is frustration-freeness --- this  represents infinite fine-tuning in the space of possible local Hamiltonians, and is connected with  a  large ground state degeneracy in the models of Ref.~\cite{watanabe2024critical}.
Separately,  Ref.~\cite{fava2024heisenberg} showed using spin-wave theory that the isotropic Heisenberg spin chain with random-sign couplings breaks SO(3) at least for large enough spin $S$.\footnote{The ${S=1/2}$ case is debated~\cite{fava2024heisenberg,li2025ground,PhysRevLett.75.4302,PhysRevB.55.12578,PhysRevB.60.12116}.}
This model is not ``frustration-free'', but it has SO(3) symmetry, rather than just U(1).
The examples above show that the simplest form of the standard lore cannot be correct.
But since they do not address   generic models with abelian continuous symmetry, 
they leave open the possibility that some more precise formulation of the lore could survive.

In this paper we study models whose only continuous symmetry is U(1). 
We argue that, contrary to the lore,  long-range order (LRO)  occurs at a zero-temperature phase transition that can be reached by tuning one parameter in  (otherwise generic) Hamiltonians with the same symmetry as the XXZ chain. 
 By a standard mapping,   this implies the possibility of superfluid order at phase transitions in boson models.
 We explain this phenomenon in terms of Berry phases in a simple field theory (imaginary terms in the Euclidean Lagrangian \cite{fradkin2013field}). These Berry phases distinguish the field theories of the 1D quantum spin systems below from those describing 2D classical spin systems, i.e. they are the reason that intuition from the Hohenberg-Mermin-Wagner theorem does not apply.

The critical point that we discuss separates a phase with conventional \textit{quasi}-long-range order\footnote{I.e. power-law correlations described by free field theory for the angle $\theta$ in the $(S^x,S^y)$ plane.} 
in $(S^x, S^y)$
from a phase where quasi-long-range order in $(S^x,S^y)$ coexists with ``Ising''--like long-range order in $S^z$.
At the critical point, $S^z$ is critical, while $(S^x, S^y)$ becomes long-range ordered, breaking U(1) symmetry.
Fig.~\ref{fig:MFphasediag} shows a schematic phase diagram topology: we focus on the transition between the two upper phases.

These phases, and the critical point between them,  can be accessed by  perturbing
 the isotropic Heisenberg Hamiltonian, i.e. via
\be\label{eq:perturbedferro}
\mathcal{H} = - \mathcal{J} \sum_{i} \vec{S}_i.\vec{S}_{i+1} + \lambda \, \mathcal{H}',
\ee
where the term $\mathcal{H}'$  breaks the SO(3) of the isotropic model down to that of the easy-plane model.
As one example of a model with this symmetry, we could take a spin-1 chain with 
$\lambda \, \mathcal{H}' = g_1 \sum_i (S^z_i)^2  + 
g_2 \sum_i (S^z_i)^2 (S^z_{i+1})^2  + \ldots$.
The analysis can be controlled by assuming that the bare parameter $\lambda$ is small
(an alternative control parameter would be large $S$).
However, the resulting  universal behavior is not dependent on the Hamiltonian being close to the isotropic limit, i.e. it should apply to ``generic'' Hamiltonians with no large or small parameters.

The analysis starts from the coherent states path integral for the spin chain.
We map the critical theory to an interacting  1+1D field theory: 
one formulation is as a Lifshitz-like field theory \cite{grinstein1981anisotropic,
fradkin2004bipartite,
bervillier2004exact,radzihovsky2011fluctuations,hsu2013dynamical} that has also arisen in the context of a critical point for 1D fermions  \cite{yang2004ferromagnetic,kozii2017ferromagnetic}.
 The critical theory can be analyzed by epsilon expansions around 2+1 dimensions or by a large $N$ expansion.
Simple lowest-order calculations in both approaches give a consistent picture for the topology of the RG flows.
Contrary to standard examples of continuous symmetry breaking,
the Goldstone mode $\delta \theta$ on top of the
ordered state,
and its  canonically conjugate field $S_z$,
remain nontrivially interacting even in the deep infra-red~(IR).

The phases discussed here have a standard translation into the language of hopping bosons,  with $S^z$ mapping to the particle density. 
Therefore in principle these phases could be studied in that setting, given appropriate interactions.
In addition to U(1) and spatial symmetries, we will usually assume a discrete symmetry that reverses the sign of $S^z$:
in the absence of such a symmetry it is necessary to tune a second parameter in order to access the universality class discussed here.\footnote{In the presence of a symmetry that reverses $S^z$, the critical point separates two symmetry-inequivalent phases. In the absence of such a symmetry the critical point is analogous to the critical endpoint of the standard liquid-gas transition, which lies within a single phase, and must be reached by tuning two parameters.}
In the boson language, this discrete symmetry exchanges particles and holes, 
and the transition we discuss is a transition between a quasi-long-range-ordered superfluid, and a phase where 
quasi-long-range  superfluidity  coexists with spontaneously broken particle-hole symmetry.\footnote{I.e. phase separation into regions of larger and smaller density.} At the (zero-temperature) critical point, the superfluid phase is long-range ordered, and the density fluctuations are critical.

\section{Perturbing the isotropic ferromagnet}

Assume for simplicity that we wish to study Hamiltonians with the same symmetry as the XXZ Heisenberg chain, but with more generic interactions.
The only continuous symmetry is 
a U(1) spin rotation symmetry around the $S^z$ axis.
This forms part of  an O(2) symmetry that also includes   $\pi$ rotations which reverse the sign of $S^z$.
Finally we have spatial translation and reflection symmetries as well as time-reversal.\footnote{An antiunitary symmetry   $\mathcal{T}$ that reverses the sign of the spin operators.}  (Some of these symmetries can be dropped without affecting the conclusions, as we will discuss later.)

Although we wish to study models with U(1) or O(2) spin symmetry, it is useful to start with the isotropic ferromagnet with full SO(3) spin rotation symmetry,
 i.e. Eq.~\ref{eq:perturbedferro} with ${\lambda=0}$.
The reason is that the continuum theory of the isotropic point is simple can be written down exactly, and  gives a starting point for adding terms that reduce spin symmetry to the desired U(1) or O(2).
The mechanism for long-range order is simplest to understand in the field theory: afterwards we will comment briefly on microscopic spin-chain Hamiltonians that realize the universal behavior of the field theory.

The imaginary-time Lagrangian for the SO(3) ferromagnet is (see e.g. Ref.~\cite{auerbach2012interacting} and App.~\ref{app:SO3pathintegral} for review)
\be\label{eq:coherentstates1}
\mathcal{S}=\int_0^\beta \dd t \int \dd x \lf
- i S (1 - \psi) \partial_t \theta 
- \f{\mathcal{J} S^2}{2} (\nabla \vec n)^2
\ri,
\ee
where ${\vec{n}=(\sqrt{1-\psi^2}\, \cos\theta,\sqrt{1-\psi^2}\,\sin\theta,\psi)}$ is the normalized spin field. The lattice spacing has been set to~1.

Since we will soon reduce the symmetry to O(2), we will often need to expand around a classical ground state with ${\psi = 0}$,
i.e. a ground state with the order parameter in the easy plane. 
Writing   $D=\mathcal{J}S$, the quadratic expansion gives\footnote{Unusually, compared to other applications of the coherent states path integral, the coefficients in Eq.~\ref{eq:initialspinwave} are exact for any size of the spin $S$, because the ground state and first excitations of the ferromagnet are given exactly by spin wave theory \cite{takahashi1987few}.} 
 \be\label{eq:initialspinwave}
\mathcal{S}_G= 
S  \int \hspace{0mm}
 \dd x \dd t\left[
 i  \, \psi \partial_t \theta
+  \f{D}{2}
\Big(
(\nabla\theta)^2 + (\nabla \psi)^2 
\Big)
\right].
\ee
$\mathcal{S}_G$ is a Gaussian RG fixed point that captures fluctuations of the Goldstone modes
(in fact since   $\theta$ and $\psi$ are conjugate, they together form a single ``type B'' Goldstone mode \cite{watanabe2020counting})
  on top of the chosen ordered states.
Power counting shows that  higher-order terms are RG--irrelevant for such fluctuations.\footnote{These terms are however important for enforcing the symmetry between ground states with \textit{different} values of $S_z$.
We have also dropped a total derivative. 
See App.~\ref{app:SO3pathintegral} for details.}
The Berry phase term, with its single time derivative, is the key difference from standard Lorentz-invariant theories, and allows the ground states of the isotropic model to show long-range order. 
That is, the reduction to the ordered moment due to spin waves is finite: there is no IR divergence. (See App.~\ref{app:SO3pathintegral} for a more detailed review.)

Now we allow for a nonzero $\lambda$ in Eq.~\ref{eq:perturbedferro}.
This will induce perturbations in Eqs.~\ref{eq:coherentstates1},~\ref{eq:initialspinwave}.
We may  
{\bf (1)} obtain the coarse structure of the resulting phase diagram by simple mean field theory, and 
 then we must {\bf (2)}  consider gapless regions of the phase diagram more carefully using RG.
If $\lambda$ is small (or if the spin size $S$ is large) 
 this process should give quantitatively accurate locations of phase boundaries.
But the universal results are robust even in the absence of a small parameter.
The critical theory we will discuss can also apply for any value of $S$, given an appropriate Hamiltonian. Note for example that $S$ can be scaled out of the Lagrangians by a coordinate rescaling.

The perturbations to the Lagrangian are constrained by the symmetries assumed above  --- O(2) symmetry,  spatial symmetries and time-reversal. The action of these symmetries is   described in App.~\ref{app:symmetryaction}.
As examples, these symmetries allow terms in $\mathcal{H}'$ of the form $\sum_i (S^z_i)^{2k}$, which are nontrivial if $S\geq k$,  
$\sum_i S^z_i S^z_{i+1} \cdots S^z_{i+2k}$, etcetera.
We wish to allow for generic combinations of such terms. 
As a conceptually simple example 
(albeit not the most natural one from the point of view of realizations)
we could imagine
a chain, with sufficiently large $S$,
with an onsite potential 
${\lambda\mathcal{H} = \sum_{k=1}^3 g_{2k} (S^z_i)^{2k}}$.
Here we have included the sextic coupling to ensure the phase diagram is fully generic, but this coupling is not required in order to access the phase transition of main interest to us.
See also Eq.~\ref{eq:modelexample} for a spin-1 example.

\begin{figure}
    \centering
\includegraphics[width=0.95\columnwidth]{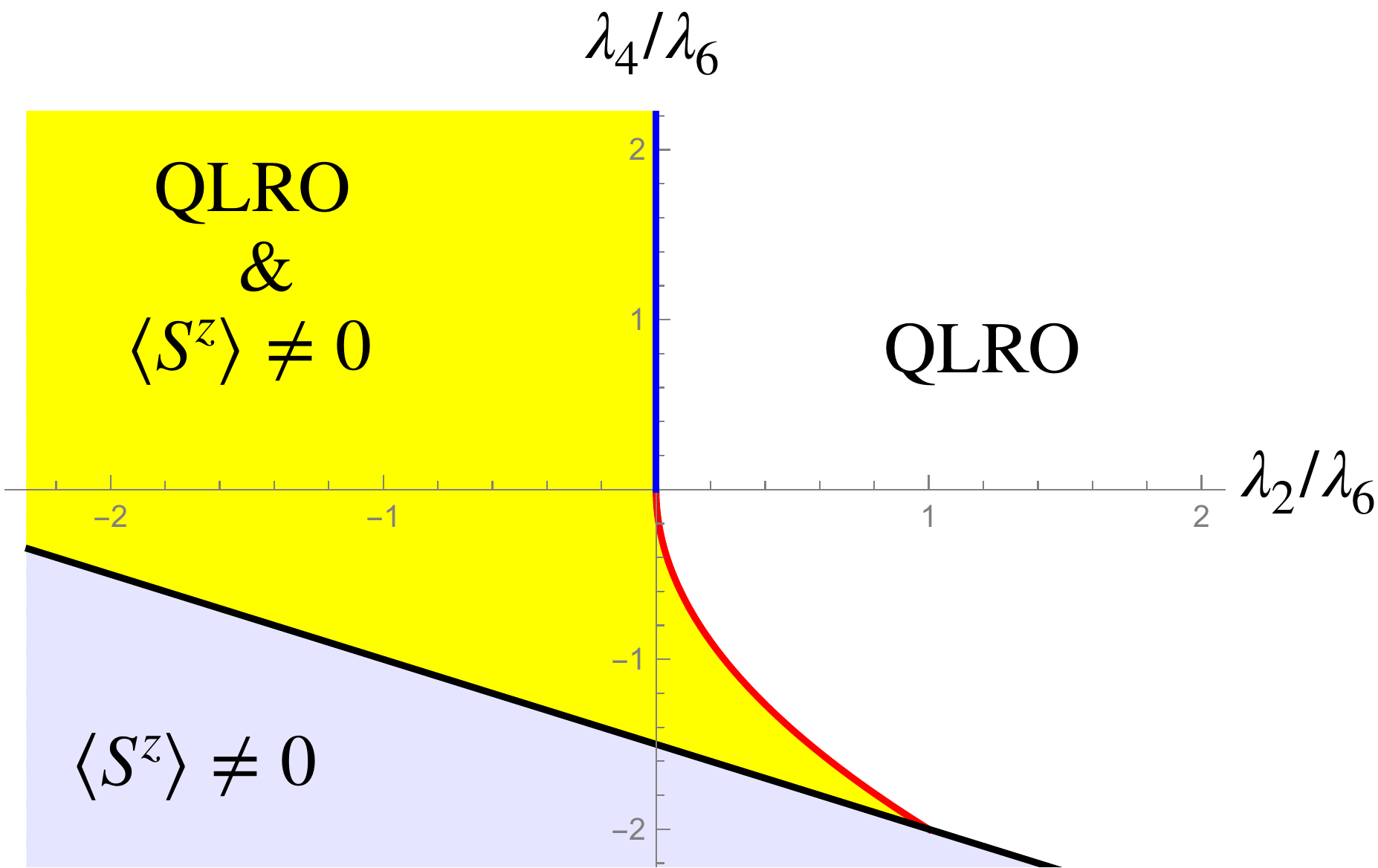}
    \caption{An illustrative mean-field phase diagram, for the case where the potential 
    for ${n_z = S^z/S}$
    in {\blue Eq.} is ${V(n_z^2) = \lambda_2 n_z^2 + \lambda_4 n_z^4 + \lambda_6 n_z^6}$ with ${\lambda_6 > 0}$. 
  Minimizing $V$ gives three  phases: 
  We can have nonzero expectation  
  values for 
  $S_z$, or for $(S_x, S_y)$, or for both.   
  Going beyond mean field converts long-range order in $(S_x,S_y)$ to quasi-long-range order (QLRO) within these phases, hence the labelling in the figure. 
  Within  mean field,
  the transitions on the lower boundary of the QLRO phase are first order while the others are continuous.
  In particular the blue (vertical) transition line (for ${\lambda_4>0}$) is continous. This transition is our main interest. 
  We argue that $(S^x, S^y)$ shows true long-range order on this line (as well as at the origin of the phase diagram).}
\label{fig:MFphasediag}
\end{figure}

Generically, such symmetry-breaking terms will induce a potential $V(\psi^2)$  
in Eq.~\ref{eq:coherentstates1}, 
which is of order $\lambda$ and given formally by 
$V(\psi^2)=
\f{\lambda}{L} 
\< \vec n| \mathcal{H}' |\vec n\>$.
(Here  $\ket{\vec{n}}$ 
 is a translationally invariant 
 product state polarized in the ${\vec n}$ direction, and $L$ is the system size.)
Step {(1)} of the procedure above is to minimize this potential.\footnote{More physically, we are picking the polarized ground state of $\mathcal{H}_{\lambda=0}$   which minimizes $\< \mathcal{H'}\>$.}
Fig.~\ref{fig:MFphasediag} shows the mean-field phase diagram for the illustrative case
\be\label{eq:meanfieldpotential}
V({ \psi^2}) = \lambda_2 \psi^2 +  \lambda_4 \psi^4 + \lambda_6 \psi^6
\ee
for (arbitrary) fixed positive $\lambda_6$.
The phase diagram with ${\lambda_6=0}$ displays all the same phases, 
and can be inferred from Fig.~\ref{fig:MFphasediag}.
Note that while the microscopic model may have many parameters, there are only two coordinates determining the phase structure in Fig.~\ref{fig:MFphasediag}. We will discuss microscopic models realizing this physics below.

This phase diagram has three phases, according to whether the minimum for $\psi^2$ is at its maximum value $1$, at its minimum value $0$, or at an intermediate value.
Within mean field theory, all three phases have LRO
--- in $\psi$, in $\theta$, or in both ---
but, as usual (reviewed below), fluctuations will replace LRO in $\theta$ with quasi-LRO within these phases, hence the labelling in the  diagram.

We now focus on the phase transition at $\lambda_4>0$ (the blue line in Fig.~\ref{fig:MFphasediag}) 
between the $\theta$--QLRO phase and the phase with both $\theta$--QLRO and  $\psi$--LRO.
We refer to this as the Q--QL transition.

\section{Critical theory}

We can go beyond mean field by including fluctuations of $\theta$ and $\psi$.
At the transition, the minimum of the mean-field potential is at ${\psi=0}$, 
so our starting point is  $\mathcal{S}_G$  in Eq.~\ref{eq:initialspinwave}.
Neglecting terms that are irrelevant by power counting,  the only new terms (not already present in $\mathcal{S}_G$) allowed by the assumed symmetries are those in Eq.~\ref{eq:meanfieldpotential}.
We expect that in order to study the universality class of the Q--QL transition it suffices to consider the Lagrangian
 \be\label{eq:Scrit}
\mathcal{L}= 
\mathcal{L}_G
+\lambda_2 \psi^2
+ \lambda_4 \psi^4,
\ee
where $\mathcal{L}_G$ includes the Gaussian derivative terms in (\ref{eq:initialspinwave}), 
potentially with modified coefficients 
(though these modifications will be small  in the UV if $\lambda$ is small).
The Q-QL transition occurs  when the (renormalized) value of $\lambda_2$ is tuned to zero.
We have omitted $\psi^6$ as it is less relevant than $\psi^4$.
After rescaling the coordinates and $\psi$,
and absorbing constants into $\lambda_2$ and $\lambda_4$,
 \be\label{eq:Lpsitheta}
\mathcal{L}'=
 i  \, \psi \partial_t \theta
+  \f{1}{2}
(\nabla\theta)^2 
+  \f{1}{2}
(\nabla \psi)^2 
 + \lambda_2 \psi^2 
+  \lambda_4 \psi^4.
\ee

Note that the two stable phases either size of the Q-QL transition 
(Fig.~\ref{fig:MFphasediag}) are recovered when $\lambda_2$ is either positive or negative. In either case, $\psi$ has massive fluctuations around a quadratic minimum: 
integrating these fluctuations out generates a $(\partial_t\theta)^2$ term, 
giving a  Lorentz-invariant theory for long-wavelength fluctuations of  $\theta$  (describing quasi-LRO).\footnote{If $|\lambda_2|$ is small, the stiffness of $\theta$ is large and its velocity is  small. }

At the phase transition line, the renormalized ${\lambda_2}$ vanishes.
Although $\mathcal{L}'$ involves two fields, they are canonically conjugate. 
Integrating $\theta$ out by the standard trick of writing ${\psi = \nabla h}$ gives an  alternative formulation with a single  field and a
Lifshitz-like 
\cite{grinstein1981anisotropic,
fradkin2004bipartite,
bervillier2004exact,yang2004ferromagnetic,radzihovsky2011fluctuations,hsu2013dynamical,kozii2017ferromagnetic}
 kinetic term (though in this form the U(1) order parameter $\theta$ is no longer explicit):
 \be\label{eq:Lcrit2}
\mathcal{L}_\text{crit}''= 
\f{1}{2} (\partial_t h)^2 
+
 \f{1}{2}
(\nabla^2 h)^2 
+\lambda_4 (\nabla h)^4.
\ee
Interestingly, this  Lifshitz Lagrangian has  appeared before, in the context of an Ising transition in itinerant fermion systems, which were  studied via  bosonization in Ref.~\cite{yang2004ferromagnetic}.\footnote{I thank J. Ruhman for making me aware of this work.} 
A connection to U(1) symmetry breaking was not made in that work.
However, Ref.~\cite{yang2004ferromagnetic}  considered an $\epsilon$ expansion for Eq.~\ref{eq:Lcrit2} and argued for flow to a nontrivial fixed point in 1+1D. (See also Ref.~\cite{kozii2017ferromagnetic}.)
This is consistent with our analysis below of the RG flow, which is based on an alternative $\epsilon$ expansion and on a large $N$ approximation.
(We will consider only lowest-order calculations, leaving higher-order  analyses to the future.)

The coupling $\lambda_4$ in Eq.~\ref{eq:Lpsitheta},
or equivalently in Eq.~\ref{eq:Lcrit2},
is RG-relevant at the Gaussian fixed point. 
We propose that (for positive $\lambda_4$)  this 1+1D theory flows to an interacting fixed point describing the Q-QL transition line,  at which LRO in $\theta$, which selects some arbitrary ordering direction $\theta_0$,
coexists with nontrivial fluctuations on top of the ordered state.

In other words
${\lim_{\beta,L\to \infty}\< [\theta(x,0) - \theta(0,0)]^2 \>}$
remains finite as $x\to\infty$,
consistent with long-range order of $e^{i\theta}$,
but the correlators of $\psi$ and of 
${\delta\theta = \theta - \theta_0}$
are characterized by nontrivial exponents.
This proposal is supported by  simple lowest-order RG calculations in the field theory, as discussed below.

Anticipating slightly, let us note some properties of the fixed point.
The Gaussian theory (${\lambda_4=0}$) has scaling dimensions ${\Delta_\theta=\Delta_\psi=1/2}$ for the two fields, and dynamical exponent $z=2$.
At the interacting fixed point these values will (a priori) change.
However the first term in (\ref{eq:Lcrit2}) cannot be renormalized 
(see App.~\ref{app:wilson} \cite{hohenberg1977theory,wiese2022theory})
and this implies $\Delta_\psi=(3-z)/2$.
The equations of motion give also ${\Delta_\theta=1-\Delta_\psi}$, so that the three exponents above reduce to a single nontrivial one.
Further, the shift symmetry ${\theta\rightarrow\theta+c}$
implies that the scaling dimension of $(\delta\theta)^k$ is just $k \Delta_\theta$.\footnote{This is necessary so that the shift in $\theta$ commutes with the RG transformation of the operators.}
(There is no such constraint for powers of $\psi$.)
Importantly, the calculations below, 
as well as more general arguments, indicate that the scaling dimension $\Delta_\theta$ is positive at the fixed point (as we would naively expect). An explicit argument ruling out a negative value of $\Delta_\theta$ is given in App.~\ref{app:DeltaThetaBound}, as well as the heuristic reason why $\Delta_\theta$ is positive.

Long-range order is stable so long as the theory indeed flows to an IR fixed point where the scaling dimension  ${\Delta_\theta}$ is 
positive.
First, the fact that ${\Delta_\theta>0}$   guarantees that  vortices in $\theta$ are RG-irrelevant and that we are justified in ignoring them in the field theory.\footnote{Note that $\Delta_\theta>0$ implies that the compactification radius of $\theta$ flows to infinity under RG (with the standard convention where the coefficients of the derivative terms are  kept fixed under RG).  Therefore the free-energy cost of a vortex increases under RG, implying that vortices are irrelevant.}
Second, flow to a  fixed point with ${\Delta_\theta>0}$
ensures that fluctuations of $\theta$ remain finite as the system size diverges.
At the fixed point, the propagator for $\delta\theta$ has the scaling form $(|k|+|\omega|^{1/z})^{2\Delta_\theta-z-1}F(w/|k|^z)$, for some scaling function $F$; integrating over momenta and frequencies then shows that the fluctuations $\<\delta \theta(x,t)^2\>$ remain finite  when ${\Delta_\theta>0}$, rather than suffering an IR divergence.
(Compare the analogous discussion of the  propagator in the free theory, Eq.~\ref{eq:initialspinwave}, in App.~\ref{app:SO3pathintegral}.)

Evidence that (\ref{eq:Lcrit2}) flows to such an IR fixed point comes  
from {\bf (1)}  an $\epsilon$-expansion around  $d=2$ spatial dimensions   for Eq.~\ref{eq:Lpsitheta}, described below,
and
{\bf (2)}  the closely-related $\epsilon$-expansion for Eq.~\ref{eq:Lcrit2} discussed in 
Ref.~\cite{yang2004ferromagnetic};\footnote{We also note  RG results for the 
3D (i.e. 2+1) Lifshitz  theory in \cite{grinstein1981anisotropic,fradkin2004bipartite,hsu2013dynamical} and closely related theories in ${3-\epsilon}$ \cite{radzihovsky2011fluctuations,kozii2017ferromagnetic}. Ref.~\cite{kozii2017ferromagnetic} discusses a variant with reduced symmetry  and makes a comparison with a smectic liquid crystal in 2D.} and finally from {\bf (3)} a large-$N$ approximation described below.

The reason that two $\epsilon$-expansions are possible is that the  theories in Eqs.~\ref{eq:Lpsitheta} and \ref{eq:Lcrit2} are equivalent in 1+1D, 
via $\psi = \nabla h$,
but no longer equivalent in higher dimensions.
Hence the 1+1D theory may be approached 
by expanding around either of the two theories in 2 space dimensions.\footnote{Directly in 2+1 dimensions, Eq.~\ref{eq:Lpsitheta} describes a transition inside a $\theta$--LRO phase. The RG equation below shows that this 2+1D transition has logarithmic corrections to mean-field scaling.}
At lowest order, the two $\epsilon$-expansions give similar flows.

Consider Eq.~\ref{eq:Lpsitheta} in general dimension $d+1$.
It is convenient to  integrate $\theta$ out, giving a $\psi^4$ theory with a nonlocal dispersion.
Explicit RG results are given in  App.~\ref{app:wilson}. After  a change of variable, the   $O(\epsilon)$ RG equations  (with ${\epsilon = 2-d}$) are just like those for isotropic  $\phi^4$ theory near its critical dimensionality
 \cite{kardar2007statistical}:
\ba
\partial_\tau  \lambda_2 & = 
 \lf
2 -  \lambda_4/3
\ri \lambda_2,
&
\partial_\tau  \lambda_4 & = 
\epsilon \lambda_4 - \lambda_4^2.
\end{align}
There is a nontrivial fixed point with a single relevant symmetry-allowed coupling.
These flows are consistent with our expected phase diagram.
If we turn on $\lambda_2$, then 
the associated crossover lengthscale $\xi \sim |\lambda_2|^{-1/y_2}$ is determined by the RG eigenvalue ${y_2 =2-\epsilon/3}$ of $\lambda_2$.
Corrections to $z$ (and $\Delta_\psi$, $\Delta_\theta$)  
away from its Gaussian value would start only at $O(\epsilon^2)$, so ${\Delta_\theta = \f{1}{2} - O(\epsilon^2)}$.

As usual, an alternative approximation is to generalize the 1+1D theory to $N$ flavors,
so that $\vec \psi$ and $\vec \theta$ become $N$-component vectors
(equivalently, $\vec h$ in Eq.~\ref{eq:Lcrit2} becomes a vector)
and to take $N$ large:
 \be\label{eq:LpsithetaN}
\mathcal{L}'=
 i  \,  \vec{\psi}. \partial_t \vec\theta
+  \f{1}{2}
(\nabla\vec \theta)^2 
+  \f{1}{2}
(\nabla \vec \psi)^2 
 + \lambda_2 \vec \psi^2
 +  \f{\lambda_4}{N} (\vec \psi^2)^2.
\ee
Eq.~\ref{eq:LpsithetaN} has a physical interpretation for any $N$, as 
a stack of $N$ Heisenberg chains  with couplings  that induce $\lambda_2$ and $\lambda_4$.\footnote{Here we wish to couple the chains via their  $(S^z)^2$ operators. When $N>1$ the O($N$) symmetry of Eq.~\ref{eq:LpsithetaN} requires fine-tuning of the microscopic couplings for the chains,  to avoid an RG-relevant O($N$)--breaking anisotropy.} This field theory is trivially solvable at ${N=\infty}$  and gives a picture qualitatively like that from the lowest-order $\epsilon$ expansion:  see App.~\ref{app:largeN} for explicit formulas. 
At the ${N=\infty}$ fixed point the exponents ${\Delta_\psi=1/2}$, ${\Delta_\theta=1/2}$ and ${z=2}$ retain their Gaussian values, while ${y_2=1}$.

The leading-order $\epsilon$ expansion and large $N$ results are  strong evidence for the existence of the nontrivial fixed point with U(1) long-range order.
Of course perturbative RG results cannot guarantee the existence of a nontrivial RG fixed point: for this simulations would be required (we will discuss these elsewhere).

We note that the \textit{multi}critical points described at the beginning of the  next Section
(at which we tune a finite number $k>1$ of parameters)
are simpler in this regard.
The remaining interactions are irrelevant at these multicritical points,  so the infra-red behavior is straightforwardly accessible and is governed simply by the free theory (App.~\ref{app:SO3pathintegral}). 
The multicritical points  therefore give  a particularly simple demonstration of our claim that U(1) LRO is possible without infinite fine-tuning.

 Finally let us comment briefly on the excitation spectrum of a finite system.
So far we have discussed fluctuations at (implicitly) nonzero momentum. 
These are described by the scaling exponents discussed above: for example the typical lowest energy  for excitations at wavevector ${\sim 1/l}$ is  of order $l^{-z}$.
However for the energy spectrum of a  finite system of  length $L$ we must also take the quantum mechanics of the zero-momentum mode into account.\footnote{Note that, as a result of rescaling during RG, the compactification radius of the zero mode $\theta_0$ is much larger than 1}
We expect this to lead to a unique ground state, and low-lying excited states with a characteristic energy scale $\propto {L^{-w} \ll L^{-z}}$.
A very heuristic coarse-graining argument suggests  ${w=2z-1}$. 
These low-lying excited states, which are built by superposing ``classical'' ground states with different values of $\theta_0$, are analogous to the ``tower of states'' in an ordered antiferromagnet~\cite{anderson1952approximate,bernu1992signature,metlitski2011entanglement}.

\section{Multicritical points}
\label{sec:multicritical}

If we are able to  independently tune two  dimensionless couplings $g_1$ and $g_2$ in our microscopic Hamiltonian, then we may be able to   tune to the multicritical point (the origin of Fig.~\ref{fig:MFphasediag})
where the field-theory couplings $\lambda_2$ and $\lambda_4$ both vanish, leaving the marginally irrelevant $\lambda_6$ interaction.
This is the multicritical point separating the continuous part of the Q-QL transition  line at $\lambda_4>0$ from the first-order segment at $\lambda_4<0$.
In principle there are still higher multicritical points at which further parameters are tuned to zero.

The existence of LRO at these multicritical points is even simpler: the  leading interactions are now RG--irrelevant or marginally irrelevant, so that in the IR the Goldstone modes are governed by the  Gaussian theory  (\ref{eq:initialspinwave}), for which straightforward  dimension counting shows that long-range order is stable (App.~\ref{app:SO3pathintegral}).

When the interactions are irrelevant, the asymptotic dispersion relation of the Goldstone modes is similar to that of the isotropic ferromagnet, except that since SO(3) symmetry is no longer present, the stiffness ``$D$'' in Eq.~\ref{eq:initialspinwave} can take different values for $\theta$ and $\psi$.
However this does {not} mean that these multicritical points are equivalent at large scales to the isotropic [SO(3)-invariant] point.  
Although the symmetry-breaking interaction terms are RG--irrelevant, they are dangerously irrelevant, 
because they have a significant effect on the global order parameter.
Whereas the isotropic ferromagnet has ground states for any value of $\<S^z\>$, these O(2)--symmetric critical models have will have a unique ground state (if $SL$ is an integer) with $\<S^z\>=0$.

An important caveat is that the above picture for the $\lambda_6$--dominated multicritical point assumes   the microscopic model to be such that the bare value of $\lambda_6$ remains nonzero (and positive) 
when $\lambda_2$ and $\lambda_4$ are tuned to zero.
While
$\lambda_6$ will remain nonzero for  sufficiently ``generic'' models  with   O(2) symmetry, many simple lattice models are 
not ``generic'' in this sense.
For example, consider   (for spin size $S\geq 1$):
\be\label{eq:modelexample}
\mathcal{H} = - \mathcal{J} \sum_{i} \vec{S}_i.\vec{S}_{i+1} + 
g_1 \sum_i (S_i^z)^2
+
g_2 \sum_i (S_i^z)^2 (S_{i+1}^z)^2.
\ee
By varying $g_1$ and $g_2$ we can vary the field-theory couplings $\lambda_1(g_1,g_2)$ and $\lambda_2(g_1,g_2)$, 
so we expect be able to access the continuous Q-QL transition at $\lambda_4>0$.
However for this model the potential $V(\psi^2)$ in Eq.~\ref{eq:meanfieldpotential}
vanishes identically at  $(g_1,g_2)=(0,0)$, returning us to the isotropic Heisenberg model. 
The phase diagram topology  of this model can be obtained from Fig.~\ref{fig:MFphasediag} by taking the limit $\lambda_6\to 0$ in that figure
(which translates the  diagonal black line upwards to intersect the origin).\footnote{A simpler Hamiltonian than (\ref{eq:modelexample}), with the same number of dimensionless coupling constants, is the spin-1 XXZ model with an onsite anisotropy
\be\label{eq:modelexample2}
\mathcal{H} = - \mathcal{J} \sum_{i}( S^x_iS^x_{i+1}
+  S^y_iS^y_{i+1} + \Delta S^z_i S^z_{i+1} )  
 + 
g_1 \sum_i (S_i^z)^2.
\ee
However the  phase diagram of this model for small $g_1$ and ${g_0\equiv 1-\Delta}$ is less clear because the ``bare'' potential ${V(\psi)\propto \< \vec n|\delta \mathcal{H}|\vec n\>}$ at the lattice scale is quadratic, and we would have to consider possible generation of $\psi^4$ from RG involving irrelevant corrections to (\ref{eq:initialspinwave}).}

So far we have considered deformations of the isotropic Heisenberg model that induce a potential $V(\psi^2)$ for $\psi$, so that at the mean field level there are ground states only for one or a few $\psi$ values.
The models of Refs.~\cite{watanabe2024critical,ogunnaike2023unifying}
are interesting as they have  exact ground states for any value of $\psi$.
The analog of this at the level of mean field theory is a deformation of the Heisenberg Hamiltonian that explicitly breaks SO(3)$ \to $O(2), but only via modifications to the derivative terms, without introducing a potential. 
However, in general (in the absence of fine-tuning) a potential will be generated under renormalization.
In the models of Refs.~\cite{watanabe2024critical,ogunnaike2023unifying} the property of having ground states for all $\psi$ is protected by the frustration-free structure.\footnote{Alternately: these Hamiltonians are related to generators for  classical  Markov processes/quantum Lindbladian dynamics \cite{ogunnaike2023unifying} with a conserved quantity  $S^z$. The dynamics  must have an equilibrium state for any choice of the charge density, corresponding to a ground state  for any value of $S^z$.}  This should correspond to special structure  in the field theory \cite{ardonne2004topological,dai2020quantum}.

\section{Models with less symmetry}

Returning to the Q-QL transition that is our main focus, 
let us briefly discuss the symmetry requirements.
Above we assumed the full symmetry of the XXZ spin chain. 
This symmetry can be reduced somewhat without introducing further relevant terms in the Lagrangian.
(See App.~\ref{app:symmetryaction} for symmetry actions.)

For example, if we reduce O(2) symmetry to U(1) while retaining time reversal, $\mathcal{T}$, we do not open the door to any new relevant couplings in (\ref{eq:Lpsitheta}).
If we drop $\mathcal{T}$, while retaining 
O(2), then an additional term can appear that is formally relevant, 
namely $(\nabla\psi)(\nabla\theta)$.
However, we do not expect this term (if small) to qualitatively change the phase diagram.
This is because it is a redundant perturbation which can be eliminated by 
a change of variable ${\theta\rightarrow \theta+\text{const.} \times \psi}$.

If we retain only U(1) symmetry and spatial symmetries, then 
we allow odd powers of $\psi$ in the Lagrangian: $V(\psi) = \lambda_1\psi + \lambda_2 \psi^2 + \lambda_3 \psi^3 + \lambda_4 \psi^4+\cdots$.
If $\lambda_1$ is small enough compared to the other terms in the potential, it can be eliminated by a change of variable.
However $\lambda_3$ gives an additional relevant coupling. 

Therefore the universality class of the continuous transition discussed in this paper (with $\lambda_4>0$) could in principle be accessed in a model with only U(1) and spatial symmetries, if we tune two parameters 
(and if we have a model that lies in the right part of parameter space).\footnote{We assume $\lambda_5$ is irrelevant at the fixed point, as it is at lowest order in the epsilon expansion.}
The resulting theory describes the critical endpoint
of a first-order line in the $(\lambda_2, \lambda_3)$ plane, analogous to the endpoint of the liquid-gas transition line.
Instead of separating two distinct phases, 
this line  lies  within the QLRO phase: since there is no longer a symmetry that reverses $S^z$,  there is no LRO phase for $S^z$. 
The  critical endpoint should be relevant to boson hopping models without particle-hole symmetry.

\section{Conclusions}

We have proposed that easy-plane magnets can spontaneously break U(1) symmetry at zero-temperature phase transitions between quasi-long-range-ordered phases.
This conclusion is based on  a Landau-Ginsburg-Wilson analysis of perturbations to the isotropic ferromagnet.

Berry phases are part of the mechanism for long-range-order at these transitions, and they are the basic distinction between the  1+1D spin path integrals discussed here and field theories for 2D classical spin models.
(Many counterintuitive phenomena arise from Berry phases \cite{fradkin2013field,haldane1983continuum,affleck1989quantum,haldane19883,senthil2024deconfined}.)
The theories we have discussed have dynamical critical exponents $z$ that are larger than 1, but this on its own is not enough to enable LRO.
To see this, consider as an example  the Lagrangian 
$\mathcal{L}=\dot \theta^2 + (\nabla^2 \theta)^2$ for a putative Goldstone mode $\theta$. This Lagrangian has $z=2$, but does not support LRO ($\theta$ is even floppier than in the usual $z=1$ theory). 

Many modifications of the XXZ chain have been explored \cite{igarashi1989ground,tonegawa1990ground,kitazawa1996phase,hikihara2001ground,sengupta2007spin,okamoto2011ground,furukawa2012ground,kjall2013phase,ueda2020roles}.
It is interesting to ask what  the simplest or most natural spin chains are that can realize the Q-QL transition.
(We have noted one relatively simple Hamiltonian that should realize the continuous transition in Eq.~\ref{eq:modelexample}.)
We will explore this question numerically elsewhere.

So far we have considered models with a single U(1) order parameter. 
However the analysis demonstrates a mechanism for ``criticality--enabled'' LRO 
that should be more general.
It will be interesting to explore spontaneous symmetry breaking for other symmetry groups and other order parameter manifolds. Berry phases are also important in the spin-wave theory which allows for long-range order in the random-sign Heisenberg chain \cite{fava2024heisenberg}: Many models could be explored, both clean and disordered.

It will be interesting to explore the field theories of the Q-QL critical point  and its variations in more detail. 
 As noted above, more than one kind of $\epsilon$ expansion  can be applied, as well as large $N$.
In these approaches we start from a Gaussian UV fixed point for $(\theta, \psi)$.
We could also consider the flow from a different UV fixed point where $\theta$ and $\psi$ are uncoupled, 
$\theta$ is a free $z=1$ boson,
and $\psi$ is governed by a critical $\psi^4$ theory. Since the coupling 
 ${\propto i \psi  \partial_t \theta}$ (Eq.~\ref{eq:Lpsitheta}) is strongly relevant, 
 this is unlikely to give a well-controlled perturbative approach in the present case. However it may suggest generalizations of the ordering mechanism in which we couple other pairs, or even larger numbers, of conformal field theories.

Returning to lattice models, the fact that the models we have discussed have ferromagnetic interactions was important to the extent that we did not want the Berry phases to cancel out. Therefore  the antiferromagnet would not have been a useful starting point.
However, similar physics could be obtained with more complex ordering patterns.
As a simple example, ferromagnetism can be mapped to spiral order by a unitary transformation of the Hamiltonian \cite{shekhtman1992moriya,dillenschneider2007vector}.

Finally, while we have used the language of magnetism, the Q-QL critical point could also be realized in systems of itinerant bosons with density-density interactions.
Hard-core bosons on the lattice may of course be mapped to spins.
If there is a particle-hole symmetry, then the picture for the two adjacent phases should be  similar to that in the spin system
(on one side of the transition there is superfluid quasi-LRO, while, on the other side, quasi-LRO coexists with particle-hole-symmetry breaking, i.e.  separation into high and low-density regions).
Perhaps this transition could be seen in a cold atom experiment.

In principle (i.e. with enough tunability) it is  possible to realize the critical behavior even without particle-hole symmetry. 
In this case the critical point is the endpoint of a first-order line within the quasi-LRO superfluid phase (critical density fluctuations at this endpoint enable superfluid LRO).\footnote{It would also be possible to use the mechanism in this paper to construct examples of  superconductivity in 1D fermionic Hamiltonians.}

Finally, let us compare the present phenomena with other kinds of long-range order in low dimensions.

First there are systems where (unlike for spin rotation symmetry) the symmetry action involves the spatial coordinate in a nontrivial way. 
While a classical 2D crystal cannot break translational symmetry,  
it can break spatial rotation symmetry \cite{kardar2007statistical}.
Two-dimensional classical  membranes embedded in three dimensions can also break (3D) rotation symmetry \cite{nelson1987fluctuations,
aronovitz1988fluctuations,
david1988crumpling,
le1992self,
le2018anomalous,
le2025tattered}. 
(Ref.~\cite{radzihovsky2024critical} discusses related ordered states, with nontrivial, non-Gaussian Goldstone modes.)
Finally, continous ``dipolar'' symmetries, in which the symmetry transformation depends on the coordinate, can also be broken in low dimensions \cite{stahl2022spontaneous,kapustin2022hohenberg}.

Because the symmetry transformations 
in these examples depend on the spatial coordinate, the energy cost (or action cost) of a ``small'' transformation, applied to a subregion, can be much larger than for a standard internal symmetry: in the crystal, for example, a ``small'' spatial rotation moves particles by a large distance if they are far from the origin \cite{kardar2007statistical}.

Second, continous symmetries can be broken at the low-dimensional spatial boundaries of higher-dimensional critical systems \cite{sun2025boundary}. In these examples the bulk effectively mediates long-range interactions, so that no Hohenberg-Mermin-Wagner-like restriction applies.

In contrast to these examples, here we have argued for the breaking of a conventional internal U(1) symmetry, in a local 1D model,
via a mechanism involving Berry phases.

\acknowledgments{I am very grateful to J. Chalker for useful feedback on the draft, to K. Wiese for helpful discussions of RG, and to X. Cao, F. Essler, D. Kovrizhin, R. Verresen, and H. Watanabe for useful discussions.}

\begin{appendix}

\section{Coherent states path integral for SO(3) ferromagnet}
\label{app:SO3pathintegral}

Recall that an imaginary-time path integral for spins $j=1,\ldots, L$
may be constructed 
by inserting resolutions of the identity $\mathbb{I}\propto \int_{\{\vec n_j\}} \ket{\vec n_1, \ldots, \vec n_L}\bra{\vec n_1, \ldots, \vec n_L}$ into $\Tr e^{-\beta H}$, where $\ket{\vec n_1, \ldots, \vec n_L}=\bigotimes_{j=1}^L\ket{\vec n_j}$, and $\ket{\vec n}$ is a spin state polarized in the direction of the unit vector $\vec n$.
For the SO(3) ferromagnet this gives an action with a crucial Berry phase term  \cite{auerbach2012interacting}:
 \be\label{eq:initialaction}
\mathcal{S}= 
 \int_0^\beta \dd t
 \sum_{j=1}^L
 \lf
- i S    (1-\psi_j)
\, \partial_t \theta_j
- 
\mathcal{J} S^2    \, \vec n_j \cdot \vec n_{j+1}
\ri.
\ee
We take periodic boundary conditions in space and imaginary time.
Expanding to second order in gradients,
\be
\mathcal{S}=\int_0^\beta \dd t \int \dd x \lf
- i S (1 - \psi) \partial_t \theta 
- \f{\mathcal{J} S^2}{2} (\nabla \vec n)^2 
\ri,
\ee
or equivalently
 \ba
 \notag
\mathcal{S}  = &  
\int \dd x \dd t\Big[ 
-i S  \,  (1-\psi) \partial_t \theta
\\ \label{eq:Sappendixnonlinear}
& \quad\quad\quad \quad +  \f{\mathcal{J}S^2}{2}
\lf
\f{(\nabla \psi)^2}{(1-\psi^2)}
+ (1-\psi^2) (\nabla\theta)^2
\ri
\Big].
\end{align}
Note that this action includes a total derivative term proportional to $\partial_t\theta$.
Since $\theta$ only has to be time-periodic modulo $2\pi$, this term is not necessarily zero. It contributes
(we return to the lattice for a moment)
\be\label{eq:Swinding}
\mathcal{S}_{\text{total deriv.}} = -i 2\pi S \int \sum_x  \, m_x, 
\ee
where $m_x$ is an integer counting the temporal winding of $\theta$ at  site $x$.
If we assume the fluctuations of $\theta$ to be small, then   $\theta(x,t)$ does not contain spacetime vortex singularities, which implies that $m_x = m$ is independent of the site, so that  
\be
\mathcal{S}_{\text{total deriv.}} = -i 2\pi S L \, m. 
\ee
Since this term depends only on the spatial zero mode of $\theta(x,t)$, it plays no role in  renormalization-group flows, and so can be discarded for our purposes.\footnote{It does play a role in ensuring that the quantization of $S^z_\text{total}$ is respected, see e.g. App.~A of \cite{fava2024heisenberg}.}  
If we take vortices into account, then $\mathcal{S}_{\text{total deriv.}}$ plays a role in determining their fugacity: for completeness we discuss this below. However, vortices will be RG-irrelevant in the phases and at the phase transitions\footnote{Assuming that $\theta$ has a positive scaling dimension $\Delta_\theta$, then  (in the usual Wilsonian RG convention where we keep the coefficient of the derivative terms in the action fixed) the compactification radius of $\theta$ is growing under RG, meaning that vortices are becoming more costly at larger scales.}  we discuss.

The  action in (\ref{eq:Sappendixnonlinear}) has terms of all orders in $\psi$.
As usual in nonlinear sigma models, the higher-order terms do not have independent coupling constants, since they are fixed by symmetry.
If we wish to discuss arbitrary ground states, these higher-order terms should be kept, so that SO(3) symmetry is respected.
However if we are interested in fluctuations on top of   ground states that lie in the XY plane (or if we are going to perturb the model with a potential whose minima are at $\psi=0$)  then we can retain only the quadratic terms in the fields. 
This is because the higher-order terms are strongly RG-irrelevant:  the engineering dimensions of the fields are
\be
\theta\sim \psi \sim \text{length}^{-1/2},
\ee
so that the RG eigenvalue of the leading nonlinear term,  $\psi^2 [(\nabla\theta)^2 - (\nabla\psi)^2]$, is $y=-1$.

The resulting quadratic theory ($D=\mathcal{J}S$)
 \be\label{eq:initialspinwaveapp}
\mathcal{S}= S
\int \dd x \dd t\left[
 i  \, \psi \partial_t \theta
+  \f{D}{2}
\Big(
(\nabla\theta)^2 + (\nabla \psi)^2 
\Big)
\right]
\ee
has trivial \textit{equal}-time correlators (as expected from the fact that the isotropic model possesses a trivial polarized ground state) but has nontrivial nonequal-time correlators. 

Formally, a simple way to see the presence of long-range order is 
from Wilsonian RG, which amounts to a trivial rescaling (since the theory is noninteracting). If we make  a coarse-graining transformation with $x\to x' b$ and $t\to t' b^2$, but do not rescale $\theta$ or $\psi$, then the diffusion constant $D$ remains invariant but the overall factor outside the action increases as $S\to bS$. Formally this means that the ``stiffness'' of this ``sigma model'' is increasing under RG, signalling long-range order. More simply, the spins are adding up coherently in the block, since the ground states are (locally) perfectly polarized.

More concretely, it is straightforward to check that the propagator for  Eq.~\ref{eq:initialspinwaveapp} is consistent with long-range order. 
In Fourier space,
$\<\theta(k,\omega) \theta(k',\omega')\>=(2\pi)^2{\delta(k+k')\delta(w+w')}
\f{D k^2}{w^2 + D^2 k^4}$.
Integrating over frequencies and momenta (bearing in mind the presence of a UV cutoff) shows that the local fluctuations\footnote{Of course we have to exclude the zero mode, which is the global order parameter}
$\<\theta(x,t)^2\>$ 
are finite: 
there is no  IR divergence.
This may be contrasted with the case of a Lorentz-invariant Lagrangian $\propto (\partial_t\theta)^2 + (\nabla\theta)^2$, for which there is a logarithmic divergence in $\<\theta(x,t)^2\>$ arising from small ${w^2 + k^2}$.

When the coherent states approach is applied to generic Hamiltonians,  the quantitative values of the couplings in the continuum Lagrangian are under quantitative control only at large $S$.\footnote{See Ref.~\cite{wilson2011breakdown} for a discussion of subtleties in the coherent states path integral.}  But for the isotropic ferromagnet, where the ground states and lowest-energy excitations are given exactly by quadratic spin wave theory, the constants in (\ref{eq:initialspinwaveapp}) are exact even for $S=1/2$. (In any case, our results are  based on RG, so do not depend on the specific numerical values of the continuum couplings.)

If we allow for vortices in $\theta$, then $S_\text{total deriv.}$ plays a role in determing the vortex fugacity, as  can be seen by a standard kind of argument (see e.g. \cite{haldane19883,read1989valence,senthil2005deconfined} for the 2D case). 
In the ferromagnet vortices are anyway strongly RG irrelevant,
but for completeness we now describe the formal effect of $S_\text{total deriv.}$.

Assume that ${m_0=0}$.  Vortices are naturally associated with bonds of the lattice.
If we place a vortex at location $(x_1+1/2, t_1)$ and an antivortex at $(x_2+1/2, t_2)$, with $0\leq x_1<x_2<L$, then $m_x$ is equal to 1 for $x\in \{ x_1 + 1, x_1 + 2, \ldots, x_2\}$, and zero elsewhere, so that, by Eq.~\ref{eq:Swinding},
\be\label{eq:vortexaction}
\mathcal{S}_{\text{total deriv.}} = -i 2\pi S (x_2 - x_1). 
\ee
If $S$ is an integer, the phase in (\ref{eq:vortexaction}) equivalent to zero, 
and so does not vary when vortices are translated. This can be interpreted as the statement that the microscopic fugacity 
(cost in the path integral) for vortices 
 is independent of which bond they occur on.
On the other hand if $S$ is half-odd-integer, then (\ref{eq:vortexaction}) shows that translating one of the vortices by a single lattice spacing changes the phase of the action by $\pi$, meaning that an elementary vortex at bond $x+1/2$ has a microscopic fugacity proportional to $(-1)^x$. 
Double-strength vortices incur twice the phase, so have $x$-independent fugacities for all $S$.

This is for the case where $\<\psi\>=0$. If we expand around the manifold of ground states with a nonzero expectation value for $\psi$, then the prefactor of (\ref{eq:Swinding}) will be proportional to ${1-\<\psi\>}$. This will modify the phase in (\ref{eq:vortexaction}) and in turn the $x$-dependence of the vortex fugacity. For a generic value of $\<\psi\>$ the vortex fugacity will oscillate (in $x$) at some wavevector $k_v$ that is incommensurate with the lattice spacing. 

The usual assumption is that vortices will only survive in the coarse-grained theory if $k_v=0$; otherwise, the coarse-grained fugacity will vanish by phase cancellation \cite{senthil2005deconfined}. For $\<\psi\>=0$, this allows single vortices for $S$ integer, and double vortices  for $S$ half-odd-integer.
For generic values of $\<\psi\>$ which yield an incommensurate $k_v$, no vortices are allowed.

\section{Symmetry actions}
\label{app:symmetryaction}

Here we describe the action of various possible symmetries on the coherent states Lagrangian.

First note that Hermiticity of the Hamiltonian implies (for the Euclidean path integral with weight $e^{-\int\mathcal{L}}$) that  
\ba
\text{Hermiticity}  & : & 
\mathcal{L}[\theta, n_z, \partial_t, \partial_x] &  = 
\mathcal{L}[\theta,  n_z, - \partial_t, \partial_x]^* .
\end{align}
In this and  the following formulas we regard $\mathcal{L}$ as an expression in the fields and derivatives, and the equalities are assumed to hold up to possible total derivative terms.
The  symmetries discussed in the main text impose:
\ba
\mathrm{U}(1)& :
&  
\mathcal{L}[\theta, n_z, \partial_t, \partial_x] 
&
= \mathcal{L}[\theta + c, n_z, \partial_t, \partial_x]
\\
\mathcal{F} & : 
&
\mathcal{L}[\theta, n_z, \partial_t, \partial_x] 
& 
= \mathcal{L}[-\theta, -n_z, \partial_t, \partial_x]
\\
\text{parity} & : 
&
\mathcal{L}[\theta, n_z, \partial_t, \partial_x] 
& 
= \mathcal{L}[\theta, n_z, \partial_t, - \partial_x]
\\
\mathcal{T}  & : & 
\mathcal{L}[\theta, n_z, \partial_t, \partial_x] &  = 
\mathcal{L}[\theta+\pi, - n_z, -\partial_t, \partial_x] .
\label{eq:Taction}
\end{align}
Here $\mathcal{F}$ is the $\pi$ rotation
${(S^x, S^y, S^z)\rightarrow (S^x, -S^y, -S^z)}$.
Time reversal $\mathcal{T}$ is an antiunitary symmetry 
($\mathcal{T}^{-1} \sqrt{-1} \mathcal{T}=-\sqrt{-1}$,
$ \mathcal{T}^{-1} \mathcal{H} \mathcal{T}=\mathcal{H}$)
that acts on spin operators as $\mathcal{T}^{-1} \vec{S}_i \mathcal{T}= - \vec{S}_i$.
(Hermiticity of $\mathcal{H}$ has been used in writing Eq.~\ref{eq:Taction}.)

Note that combinations of the above symmetries may survive even when the above symmetry generators are broken. 
For example if we add a uniform Z-field to the XXZ Hamiltonian, 
\be
\mathcal{H} = - \mathcal{J} \sum_i \lf 
S^x_i. S^x_{i+1} + S^y_i. S^y_{i+1} + \Delta
 S^z_i. S^z_{i+1} \ri - h \sum_i S^z_i,
\ee
then both $\mathcal{F}$ and $\mathcal{T}$ are explicitly broken by the field. But a modified time-reversal symmetry $\mathcal{T}'$ is retained (which is a product of $\mathcal{T}$ with a $\pi$ spin rotation). 
We can take $\mathcal{T}'$ to act simply by complex conjugation in the $S^z$ basis, since the Hamiltonian above is real in this basis. $\mathcal{T}'$ reverses the sign of $S^y$ but not of $S^x$ or $S^z$.
This symmetry imposes:
\ba
\mathcal{T}'  & : & 
\mathcal{L}[\theta, n_z, \partial_t, \partial_x] &  = 
\mathcal{L}[- \theta,  n_z, -\partial_t, \partial_x] .
\label{eq:Taction}
\end{align}

\section{$\epsilon$ expansion}
\label{app:wilson}

We take the $(\psi,\theta)$ theory in Eq.~\ref{eq:Lpsitheta}, or more generally its $N$-component generalization (\ref{eq:LpsithetaN}), in $2-\epsilon$ spatial dimensions.
Integrating $\theta$ out gives
a nonlocal Lagrangian for $\psi$ ($\hat\lambda_2 = 2 \lambda_2$)
\be
\f{1}{2} \vec{\psi} .\left[ 
\hat \lambda_2  
+  (-\nabla^2) 
+   \partial_t^2 \nabla^{-2}
\right] \vec\psi
+ \lambda_4 (\vec{\psi}^2)^2.
\ee
We impose a cutoff $|k|\leq \Lambda$ on momentum but not on frequency.
The one-loop Wilsonian RG computation is completely  parallel to that for standard $\phi^4$ theory \cite{kardar2007statistical}.
At this order only $\hat \lambda_2$ and $\lambda_4$ are renormalized.
At higher orders there will be also be a renormalization of the coefficient of $\nabla^2$, which will give a nonzero anomalous dimension and a change to $z$ which will start at $O(\epsilon^2)$.
However the coefficient of $\partial_t^2 \nabla^{-2}$ should not be renormalized at any order, as this term is nonlocal  \cite{hohenberg1977theory}. 
In 1D, the fact that the coefficient of this term is protected can also be seen in the equivalent $h$ formulation by considering the response of the free energy to a shift $h\rightarrow h + c t$, for a constant $c$ \cite{wiese2022theory}. 
I thank K. Wiese for pointing this out.
The consequence of the nonrenormalization is the relation between $z$ and the anomalous dimension noted in the text. 

To lowest nontrivial order  in $\lambda_4$,
\ba
\partial_\tau \hat \lambda_2 & = 2 \hat \lambda_2 + 4 (N+2)  \lambda_4 I_1 + O(\lambda_4^2),
\\
\partial_\tau \lambda_4  & = \epsilon \lambda_4  - (4N+32) \lambda_4 ^2 I_2 + O(\lambda_4^3),
\end{align}
similarly to the isotropic  theory \cite{kardar2007statistical}, with ($K_d = 2^{1-d}\pi^{-d/2}/\Gamma(d/2)$)
\ba
I_k = K_d
\int_{-\infty}^\infty  \f{\dd w}{2\pi} \f{1}{(\lambda_2+\Lambda^2+ w^2/\Lambda^2)^k}.
\end{align}
i.e. $I_1  = \f{K_d  \Lambda}{2 \sqrt{  (\lambda_2+ \Lambda^2) } }$,
$I_2  = \f{K_d  \Lambda}{4  (\lambda_2+ \Lambda^2)^{3/2}}$.
Expanding in $\lambda_2$ and absorbing multiplicative constants into the couplings,
\ba
\partial_\tau \hat \lambda_2 & = 
\f{2(2+N)}{(8+N)} \lambda_4 + \left[ 
2 - \f{2+N}{8+N} \lambda_4
\right]\hat  \lambda_2,
\\
\partial_\tau  \lambda_4 & = 
\epsilon \lambda_4 - \lambda_4^2 .
\end{align}
Finally, a change of variable $\hat \lambda_2 \rightarrow \hat \lambda_2 + \text{const}\times \lambda_4$ gives the RG equations in the standard form  
\ba
\partial_\tau \hat \lambda_2 & = 
 \left[ 
2 - \f{2+N}{8+N} \lambda_4
\right]\hat  \lambda_2
&
\partial_\tau  \lambda_4 & = 
\epsilon \lambda_4 - \lambda_4^2.
\end{align}
At lowest order in $\epsilon$ (in $2-\epsilon$ spatial dimensions), the RG equations coincide with those of the isotropic model in $4-\epsilon$ dimensions.

\section{Sign of $\Delta_\theta$}
\label{app:DeltaThetaBound}

In this Appendix we show that $\Delta_\theta \geq 0$ at an interacting fixed point for Eq.~\ref{eq:Lpsitheta}. 
We also give an intuitive (heuristic) explanation for why ${\Delta_\theta>0}$.

We start by assuming that  Eq.~\ref{eq:Lpsitheta} flows to a nontrivial fixed point when the renormalized $\lambda_2$ is tuned to zero. In view of the scaling relation ${\Delta_\theta = 1- \Delta_\psi}$ discussed in the text,
and the identity ${\psi  = \nabla h}$,
the desired bound  ${\Delta_\theta\geq 0}$
is equivalent to ${\Delta_h\leq 0}$. 
Therefore it is sufficient to show that assuming ${\Delta_h > 0}$ leads to a contradiction.

Recall that that $h$ is governed by the Lifshitz Lagrangian $\mathcal{L}_\text{crit}''$ in Eq.~\ref{eq:Lcrit2}.
[It is implicit that it is the renormalized coupling for $(\partial_x h)^2$, rather than the bare coupling, that is tuned to zero.]
If it was true that ${\Delta_h>0}$ for this theory, it would follow that  the correlator
\be
\label{eq:boundedcorr}
\<[h(x,t)-h(0,t)]^2\>\sim \text{const} + \text{const}.\times |x|^{-\Delta_h}
\ee
remained bounded at large $|x|$.

On the other hand, consider the perturbed Lagrangian 
\be
{\mathcal{L}_\text{crit}'' + \mu \, (\nabla h)^2}.
\ee
It is clear that for sufficiently large positive $\mu$, the  long distance correlations are then governed by a Lorentz-invariant Lagrangian proportional to ${\f{1}{v(\mu)} (\partial_t h)^2 + v(\mu) (\partial_x h)^2}$ for some $v(\mu)$.
(Indeed we  expect this to be true even for an infinitesimal perturbation, but here we avoid using  any assumption about the RG flow close to the nontrivial fixed point.)
Higher-order terms are irrelevant for this Lorentz-invariant theory. 
Therefore in the perturbed theory (which we consider for some sufficiently large fixed value of $\mu$) we have the usual logarithmic wandering at large $x$,
\be\label{eq:logfluc}
\<[h(x,t)-h(0,t)]^2\>_{\mu}\sim \ln |x|.
\ee

However the added $\mu$ term acts to suppress fluctuations of $h$.  Therefore we certainly expect  (though we do not give a rigorous proof) that at large distances 
\be
\<[h(x,t)-h(0,t)]^2\> \geq\<[h(x,t)-h(0,t)]^2\>_\mu.
\ee
Comparing with (\ref{eq:logfluc}) and 
(\ref{eq:boundedcorr}) shows that it is inconsistent to assume that ${\Delta_h>0}$, and therefore (given the scaling relation between exponents)
inconsistent to assume that ${\Delta_\theta<0}$.

Next we make the more heuristic point, by comparing the theories
\ba\label{eq:Lpsithetamulti}
\mathcal{L}_0&=
 i  \, \psi \partial_t \theta
+  \f{1}{2}
(\nabla\theta)^2 
+  \f{1}{2}
(\nabla \psi)^2 ,
\\
\mathcal{L}_\text{2}&=
 i  \, \psi \partial_t \theta
+  \f{1}{2}
(\nabla\theta)^2 
+  \f{1}{2}
(\nabla \psi)^2 
 + \lambda_2 \psi^2 ,
\\
\mathcal{L}_\text{4}&=
 i  \, \psi \partial_t \theta
+  \f{1}{2}
(\nabla\theta)^2 
+  \f{1}{2}
(\nabla \psi)^2 
+  \lambda_4 \psi^4.
\end{align}
(Again in $\mathcal{L}_4$ it is implied that it is the renormalized value of the quadratic coupling that is tuned to zero.)

Note that, because of the ``$i$'', fluctuations of $\psi$ act to \textit{suppress} fluctuations of $\theta$.
For example, compare $\mathcal{L}_0$ with $\mathcal{L}_2$.
The added mass term for $\psi$ means that fluctuations of $\psi$ are weaker for  $\mathcal{L}_2$ than for $\mathcal{L}_{0}$.
Correspondingly, fluctuations of $\theta$ are {stronger} for $\mathcal{L}_2$ than for $\mathcal{L}_0$.
In the former theory, integrating $\psi$ out leads  to a Lorentz-invariant action for $\theta$ at large scales, with ${\<(\theta - \theta)^2\>\sim \ln |x|}$, i.e. with $\Delta_\theta=0$.
By contrast in the latter theory we have $\Delta_\theta=1/2$, so that 
${\<(\theta - \theta)^2\>}$ is bounded at large separations. 

For $\mathcal{L}_4$, the fluctations of $\psi$ are 
softer than for $\mathcal{L}_2$, but not as soft as in $\mathcal{L}_0$. 
Therefore the natural expectation is that $\Delta_\theta$ lies between its values in the other two theories, i.e. ${0<\Delta_\theta<1/2}$.

\section{Large $N$ theory}
\label{app:largeN}

Consider a large--$N$ version of the theory in Eq.~\ref{eq:Lcrit2}
 \be\label{eq:LlargeN}
\mathcal{L}_{N}= 
\f{1}{2} \left[ (\partial_t \vec h)^2 
+
D (\nabla^2 \vec h)^2 
+ \lambda_2 (\nabla \vec h)^2
+ \f{\lambda_4}{2N} [(\nabla \vec h)^2]^2
\right],
\ee
where $\vec h = (h_1, \vec h_\perp)$ is an $N$-component vector. We have rescaled the coefficients for convenience.

Standard methods give the correlation functions directly in $N=\infty$ limit  \cite{cardy1996scaling}. 
In full analogy to conventional $\phi^4$ theory, the self-consistency equation
\ba
t    & = \lambda_2 + 2 \lambda_4  \int_{-\infty}^{\infty} \f{\dd w}{2\pi} \int^\Lambda_{-\Lambda} \f{\dd k}{2\pi}
\f{k^2 }{w^2 + t k^2 + D k^4}
\\
& =  \lambda_2  + 
\lambda_4\f{\sqrt{D \Lambda^2 + t}-\sqrt{t}}{D\pi}
\end{align}
determines the two-point function ($\vec \psi = \nabla\vec h$):
\be
\< \nabla h_1(x,t) \nabla h_1(0,0)\> = \int_{-\infty}^{\infty} \f{\dd w}{2\pi} \int^\Lambda_{-\Lambda} \f{\dd k}{2\pi}
\f{k^2 \, e^{i k x + i w t}}{w^2 + t k^2 + D k^4}
\ee
in the regime $t\geq 0$ (i.e. so long as we are not in the ordered phase for $\vec \psi=\nabla \vec h$).
 $t$ can be viewed as the renormalized value of $\lambda_2$, as it sets the coefficient of $k^2$ in the dispersion relation for $h_1$.
The critical point $(\lambda_2)^*$ is where ${t=0}$:
\be
(\lambda_2)^* = - \f{\Lambda}{\pi \sqrt{D}} \lambda_4.
\ee
Writing $\lambda_2 = (\lambda_2)^* + \delta \lambda_2$, the near-critical scaling of $t$ is
\be\label{eq:tlambdareln}
t \simeq \lf \f{\pi D}{\lambda_4} \ri^2 \delta \lambda_2^2.
\ee
$t$ sets a lengthscale $\xi$ above which the correlation function crosses over to the behavior of the $t>0$ phase, which is (for equal times, and neglecting oscillatory terms arising from the hard cutoff)
\be
\< \nabla h_1(x,0) \nabla h_1(0,0)\> \simeq \f{-1}{2\pi \sqrt{t}} \f{1}{x^2}.
\ee
The crossover lengthscale for this behavior is
(up to an order 1 constant)
 $\xi \propto \sqrt{D/t}$
or
\be
\xi \propto \f{\lambda_4}{\pi \sqrt{D} \, \delta \lambda_2}.
\ee
So if we write $\xi \propto \delta \lambda_2^{-\nu}$ then $\nu = 1$. This differs from the mean-field value $\nu^{\mathrm MF}=1/2$, as a result of (\ref{eq:tlambdareln}).
As in the O($N$) model we infer that the scaling dimension of $( \nabla \vec h)^2$  is  $\Delta_{(\nabla\vec h)^2}=2$ rather than the engineering value  $\Delta_{(\nabla\vec h)^2}^{\mathrm MF}=1$.
Again as in the O($N$)  model, $\nabla \vec h$ retains its engineering scaling dimension at large $N$.

If we increase the spatial dimension we can choose to consider either the $\vec h$ theory in Eq.~\ref{eq:LlargeN} or the $(\vec \theta, \vec \psi)$ theory in Eq.~\ref{eq:LpsithetaN}, since these are equivalent in $d=1$ but not in higher $d$. However at $N=\infty$  they obey the same self-consistency equation. Taking the latter theory for concreteness, in $d=2$,
\ba\notag
t    & = \lambda_2 + 2 \lambda_4  \int_{-\infty}^{\infty} \f{\dd w}{2\pi} \int_{|k|<\Lambda} \f{\dd^2 k}{(2\pi)^2}
\f{k^2 }{w^2 + t k^2 + D k^4}
\\
& =  \lambda_2  + 
\lambda_4 \lf \f{\Lambda^2}{4 \pi D^{1/2}} + \f{t \ln t}{8 \pi D^{3/2}} + \ldots \ri,
\end{align}
So that $(\lambda_2)^* = -  \f{\Lambda^2 }{4 \pi \sqrt{D}} \lambda_4$
and $t \simeq  \f{8\pi D^{3/2}}{\lambda_4} \f{\delta\lambda_2}{\ln 1/\delta \lambda_2}$.

\end{appendix}

\bibliographystyle{apsrev4-1}
\bibliography{refs.bib}
\end{document}